\documentclass{aims}
\usepackage{amsmath}
\usepackage{paralist}
\usepackage{graphics} %% add this and next lines if pictures should be in esp format
\usepackage{epsfig} %For pictures: screened artwork should be set up with an 85 or 100 line screen
\usepackage{graphicx}
\usepackage{epstopdf} %This is to transfer .eps figure to .pdf figure; please compile your paper using PDFLeTex or PDFTeXify.
\usepackage[colorlinks=true]{hyperref}
\hypersetup{urlcolor=blue, citecolor=red}

\def\currentvolume{X}
 \def\currentissue{}
  \def\currentyear{2016}
   \def\currentmonth{}
    \def\ppages{}
     \def\DOI{}

\theoremstyle{definition}

\usepackage{bm}
\newcommand{\re}{\mathrm{e}}

\newcommand{\rd}{\mathrm{d}}
\title[Acoustics in Green--Naghdi Barotropic Fluids] %Use the shortened version of the full title
      {Nonlinear Acoustics and Shock Formation in Lossless Barotropic Green--Naghdi Fluids}

\author[Ivan C. Christov]{}


\email{christov@purdue.edu}

\begin{document}
\maketitle

% Enter the first author's name and address:
\centerline{\scshape Ivan C. Christov}%$^*$
\medskip
{\footnotesize
 \centerline{School of Mechanical Engineering, Purdue University}
 \centerline{West Lafayette, IN 47907, USA}
}

%\bigskip

% The name of the associate editor will be entered by an editorial staff
% "Communicated by the associate editor name" is not needed for special issue.
% \centerline{(Communicated by the associate editor name)}

%The abstract of your paper
\begin{abstract}
The equations of motion of lossless compressible nonclassical fluids under the so-called Green--Naghdi theory are considered for two classes of barotropic fluids: (\textit{i}) perfect gases and (\textit{ii}) liquids obeying a quadratic equation of state. An exact reduction in terms of a scalar acoustic potential and the (scalar) thermal displacement is achieved. Properties and simplifications of these model nonlinear acoustic equations for unidirectional flows are noted. Specifically, the  requirement that the governing system of equations for such flows remain hyperbolic is shown to lead to restrictions on the physical parameters and/or applicability of the model. A weakly nonlinear model is proposed on the basis of neglecting only terms proportional to the square of the Mach number in the governing equations, without any further approximation or modification of the nonlinear terms. Shock formation via acceleration wave blow up is studied numerically in a one-dimensional context using a high-resolution Godunov-type finite-volume scheme, thereby verifying prior analytical results on the blow up time and contrasting these results with the corresponding ones for classical (Euler, \textit{i.e.}, lossless compressible) fluids.
\end{abstract}

% ----------------------------------------------------------------
\section{Introduction}\label{sec:intro}

Recently, there has been significant interest in the mathematics of nonlinear acoustics \cite{K15} and, specifically, in proving abstract mathematical results for model nonlinear acoustic equations, including well-posedness and control \cite{KL09,K10,KLV11,KL12,BK14,BKR14,B16}. To this end, it is important to examine the model equations and understand their origin and applicability, in order for the mathematical results to have clear implications for a diverse set of physical contexts \cite{BJ15}. Broadly speaking, model equations of nonlinear acoustics \cite{C79,mh97,j16} are derived with the goal of including only those physical effects important to a given application, while discarding physical effects deemed secondary or unimportant.

In the present work, we focus on a class of models for nonlinear acoustics in lossless fluids that conduct heat, building on a general theory \cite{gn95,gn96} based on abstract considerations \cite{gn95i,gn95ii,gn95iii} proposed by Green and Naghdi (GN) two decades ago.\footnote{Here, we do not advocate for the GN theory. Indeed it has its drawbacks; criticisms can be found in the Mathematical Reviews entry for \cite{gn95i,gn95ii,gn95iii}, while open problems are noted in \cite{b13}. In the present work, we are merely interested in understanding some of the acoustics implications of a GN theory of lossless compressible fluids that conduct heat.} Since Green and Naghdi's initial work, there has been interest in understanding both acoustic (``first sound'') \cite{js06,QS08,bs08} and thermal (``second sound'') \cite{bs08,bsj08,S10,QS12,S11,b13} nonlinear wave propagation under such \emph{nonclassical} continuum theories, which motivates the present study.

Although we consider only the lossless case in this work, viscous and thermoviscous acoustic wave propagation under the classical continuum theories (\textit{e.g.}, Navier--Stokes--Fourier) \cite{mo96,mo97,mo97b,s00,j04,ctw09,rsgc11,jncw12,db12,j13,j14,cjcw16,rsgc16}, as well as in inviscid but thermally relaxing gases \cite{j14a}, has also received significant attention in the literature.

Specifically, in Section~\ref{sec:gov_eq} of the present work, we consider the governing equations of motion of a lossless compressible Green--Naghdi (GN) fluid that conducts heat. The equations of motion are made dimensionless in Section~\ref{sec:nd}. Then, in Section~\ref{sec:baro}, the discussion is specialized to two specific classes of barotropic fluids (\textit{i.e.}, ones for which the thermodynamic pressure is a function of the density alone): (\textit{i}) perfect gases and (\textit{ii}) liquids obeying a quadratic equation of state. In Section~\ref{sec:model_eqs}, an exact reduction to a coupled system of model nonlinear acoustic equations is achieved, using a scalar acoustic potential and the (scalar) thermal displacement, and these equations' properties and simplifications are also noted. Additionally, in Section~\ref{sec:wn_eqs}, \emph{weakly} nonlinear acoustic equations for lossless GN fluids are proposed on the basis of neglecting only terms proportional to the square of the Mach number in the governing equations, without any further approximation or modification of the nonlinear terms. Then, in Section~\ref{sec:shocks}, shock formation via acceleration wave blow up is studied numerically in a one-dimensional context using a high-resolution Godunov-type finite-volume scheme. In particular, prior exact results on the blow up time are verified and contrasted with the corresponding ones for classical (Euler) fluids. Finally, Section~\ref{sec:conc} summarizes our results and notes potential future work.

% ----------------------------------------------------------------
\section{Governing equations}\label{sec:gov_eq}

A lossless compressible GN fluid \cite{gn95} in homentropic flow [Eq.~\eqref{eq:dim_entropy}] is described by the Euler equations [Eqs.~\eqref{eq:dim_mass_cons} and \eqref{eq:dim_mom_cons}] of inviscid compressible hydrodynamics \cite{th72,s97}, which are augmented by a term involving the gradient of the thermal displacement from GN theory \cite{js06} and coupled to an energy equation involving the latter [Eq.~\eqref{eq:dim_therm_disp}]:
\begin{subequations}\begin{align}
\varrho_t + \bm{\nabla}\bm{\cdot}(\varrho \bm{v}) &= 0,
\label{eq:dim_mass_cons}\\
\varrho\left[\bm{v}_t +\tfrac{1}{2}\bm{\nabla} |\bm{v}|^2 - \bm{v} \bm{\times} (\bm{\nabla} \bm{\times} \bm{v})\right] &= -\bm{\nabla}\wp -\tfrac{1}{2} m \varrho \bm{\nabla}|\bm{\delta}|^2,\label{eq:dim_mom_cons}\\
\bm{\nabla}\bm{\cdot}(\varrho\bm{\delta}) &= 0,\label{eq:dim_therm_disp}\\
\dot{\eta} = \bm{\nabla}\eta &= 0.\label{eq:dim_entropy}
\end{align}\label{eq:dim_cons}\end{subequations}
Here, $\bm{v}(\bm{x},t)$ is the velocity vector, $\varrho(\bm{x},t)(>0)$ is the mass density, $\wp(\bm{x},t)(>0)$ is the thermodynamic pressure, $\eta(\bm{x},t)$ is the specific entropy, $\bm{\delta}(\bm{x},t)$ is the gradient of the thermal displacement from GN theory,\footnote{The thermal displacement $\alpha$ is defined to be such that the absolute temperature $\theta(\bm{x},t) = \dot\alpha(\bm{x},t)$, where a superimposed dot denotes the material derivative \cite{gn95,js06}; hence, $\bm{\delta} = \bm{\nabla}\alpha$.} $m$ is the (constant) ``GN parameter'' \cite{js06}, which carries units of $(\mathsf{L}/\mathsf{T})^4/\mathrm{K}^2$ with $\mathsf{L}$ and $\mathsf{T}$ denoting length and time, respectively, all body forces have been omitted, and $t$ subscripts denote partial differentiation with respect to time $t$. To reiterate: the contributions of GN theory manifest themselves in the form of a (new) term on the right-hand side of the momentum equation \eqref{eq:dim_mom_cons} and the additional solenoidality condition \eqref{eq:dim_therm_disp}.

We assume that, initially, the fluid is in its equilibrium state, \textit{i.e.},  $\varrho=\varrho_0$, $\bm{v}=\bm{0}$, $\wp=\wp_0$, $\bm{\delta} = \bm{\delta}_0$ and $\eta=\eta_0$ at time $t=0$, where all the quantities demarcated with the subscript ``$0$'' are constant. Finally, we henceforth restrict to longitudinal (\textit{i.e.}, acoustic) waves, which implies irrotational flows, namely $\bm{\nabla}\bm{\times}\bm{v}=\bm{0}$ at the initial instant of time $t=0$ and, hence, for all $t>0$ \cite{th72}. The system~\eqref{eq:dim_cons} is not closed until $\wp$ is specified in terms of the other thermodynamic variables through an equation of state. In Section~\ref{sec:baro}, we introduce the two classes of \emph{barotropic} (meaning  $\wp = \wp(\varrho)$ only \cite{th72,OO04}) equations of state that we restrict our discussion to.

\subsection{Unidirectional flow}\label{sec:1d}

In this section, we consider the unidirectional flow of a lossless barotorpic GN fluid along the $x$-axis, which renders the problem one-dimensional (1D). Specifically, we take $\bm{v} = (v(x,t),0,0)$ and $\bm{\delta} = (\delta(x,t),0,0)$. Then, Eqs.~\eqref{eq:dim_cons} take the form
\begin{subequations}\begin{align}
\varrho_t + (\varrho v)_x &= 0,
\label{eq:dim_mass_cons_1d}\\
\varrho\left[{v}_t +\tfrac{1}{2}({v}^2)_x\right] + \wp_x &= - \tfrac{1}{2}\varrho m ({\delta}^2)_x,\label{eq:dim_mom_cons_1d}\\
(\varrho{\delta})_x &= 0,\label{eq:dim_delta_cons_1d}
\end{align}\label{eq:dim_cons_1d}\end{subequations}
where $x$ and $t$ subscripts denote partial differentiation with respect to $x$ and $t$, respectively.

Integrating Eq.~\eqref{eq:dim_delta_cons_1d} and enforcing the equilibrium condition, namely $\varrho = \varrho_0$ and $\delta=\delta_0$ at $t=0$, connects the thermal displacement gradient to the fluid's density:
\begin{equation}
\delta(x,t) = \varrho_0\delta_0 /\varrho(x,t).
\label{eq:delta-rho}
\end{equation}
Then, we can rewrite the 1D momentum equation \eqref{eq:dim_mom_cons_1d} by adding Eq.~\eqref{eq:dim_mass_cons_1d} to it and employing the result from Eq.~\eqref{eq:delta-rho}:
\begin{equation}
(\varrho {v})_t + \left(\varrho {v}^2 + \wp + m \varrho_0^2\delta_0^2 \varrho^{-1}\right)_x =  0.
\label{eq:dim_mom_cons_1d_2}
\end{equation}
Clearly, in this 1D context, the (gradient of the) thermal displacement appears as an additive contribution to the thermodynamic pressure $\wp$. Specifically, Eq.~\eqref{eq:dim_mom_cons_1d_2} is precisely the 1D Euler momentum equation for a barotropic fluid, if we define the effective pressure $\tilde{\wp}(\varrho) :=  \wp(\varrho) + m \varrho_0^2\delta_0^2 \varrho^{-1}$.

% ----------------------------------------------------------------
\section{Nondimensionalization}\label{sec:nd}

Let us now rewrite the dimensional variables using the following dimensionless ones denoted by a $\star$ superscript:
\begin{multline}
\bm{x} = \{x,y,z\} = L\{x^\star,y^\star,z^\star\} = L\bm{x}^\star,\qquad t = (L/c_0)t^\star,\qquad %\phi = (u_{\mathrm{m}}L)\phi',\\
\wp = (\varrho_0c_0^2)\wp^\star,\\ \varrho = \varrho_0\varrho^\star,\qquad \bm{v} = v_{\mathrm{m}}\bm{v}^\star,\qquad \bm{\delta} = \delta_0 \bm{\delta}^\star,
\label{eq:ndv}
\end{multline}
where the positive constants $v_{\mathrm{m}} := \max_{\bm{x},t}|\bm{v}|$, $L$ and $\delta_0 := |\bm{\delta}_0| \equiv \sqrt{\bm{\delta}_0\bm{\cdot}\bm{\delta}_0}$ denote a characteristic speed, length and thermal displacement gradient, respectively, and $c_0$ is the sound speed in the undisturbed fluid.\footnote{In the case of perfect gases, for example, $c_0 =\sqrt{\gamma\wp_0/\varrho_0}$, where the constant $\gamma(>1)$ is known as \emph{the adiabatic index}~\cite{th72,mo96,b97,s97}, which furnishes the scaling $\wp_0 \propto \varrho_0 c_0^2$ used in the nondimensionalization scheme in Eq.~\eqref{eq:ndv}.}  In terms of the dimensionless variables in Eq.~\eqref{eq:ndv}, and recalling that we have assumed an irrotational flow, the governing system of equations \eqref{eq:dim_cons} becomes
\begin{subequations}\begin{align}
\varrho_t + \epsilon \bm{\nabla}\bm{\cdot}(\varrho \bm{v}) &= 0,
\label{eq:mass_cons_nd}\\
\varrho\left(\bm{v}_t +\tfrac{1}{2}\epsilon\bm{\nabla} |\bm{v}|^2\right) &= -\epsilon^{-1}\bm{\nabla}\wp -\tfrac{1}{2}\varsigma^2 \varrho \bm{\nabla}|\bm{\delta}|^2,
\label{eq:mom_cons_nd}\\
\bm{\nabla}\bm{\cdot}(\varrho\bm{\delta}) &= 0,\label{eq:delta_cons_nd}
\end{align}\label{eq:cons_nd}\end{subequations}
where we have left the $\star$ superscripts understood without fear of confusion. %since, henceforth, we only use the dimensionless variables.
Here, we have introduced two dimensionless groups: $\epsilon := v_{\mathrm{m}}/c_0$ is the Mach number,  and $\varsigma^2 := (m \delta_0^2)/(c_0 v_\mathrm{m})$ could be termed the ``GN number'' (squared), a quantity measuring the relative effects of the thermal displacement field to the acoustic field. Of course, in the limit of $\varsigma \to 0^+$, Eqs.~\eqref{eq:mass_cons_nd} and \eqref{eq:mom_cons_nd} reduce to the dimensionless Euler equations \cite{th72,s97}.

% Note that Jordan and Straughan \cite{js06} used a reference temperature $\theta^\star$ to make the thermal displacement potential dimensionless, therefore, in \cite{js06}, $\delta_0 = c_0/\theta^\star$ and $\tilde\varsigma^2 = (m\delta_0^2)/c_0^2$, which is related to $\varsigma^2$ as $\tilde\varsigma^2 = \epsilon \varsigma^2$.

\subsection{Unidirectional flow}\label{sec:1d_sys}

Similarly, the 1D system of equations \eqref{eq:dim_cons_1d} can be made dimensionless using the variables in Eq.~\eqref{eq:ndv}. Then, upon replacing Eq.~\eqref{eq:dim_mom_cons_1d} with Eq.~\eqref{eq:dim_mom_cons_1d_2}, Eqs.~\eqref{eq:dim_cons_1d} become
%\begin{subequations}\begin{align}
%\frac{\varrho_0}{L/c_0}\varrho'_{t'} + \frac{\varrho_0 u_\mathrm{m}}{L}(\varrho' v')_{x'} &= 0,\\
%\frac{\varrho_0 u_\mathrm{m}}{L/c_0}(\varrho' {v'})_{t'} + \left(\frac{\varrho_0 u_\mathrm{m}^2}{L}\varrho' {v'}^2 + \frac{\varrho_0 c_0^2}{L}\wp' - \frac{m \varrho_0^2\delta_0^2}{\varrho_0 L}{\varrho'}^{-1}\right)_{x'} &=  0.
%\end{align}\end{subequations}
%or, leaving the primes understood,
\begin{subequations}\begin{align}
 \varrho_{t} + (\epsilon\varrho v)_{x} &= 0,
\label{eq:mass_cons_1d_nd}\\
 (\varrho {v})_{t} + \left[\epsilon\varrho {v}^2 + \epsilon^{-1}(\wp + \epsilon\varsigma^2{\varrho}^{-1})\right]_{x} &= 0,\label{eq:mom_cons_1d_nd}
\end{align}\label{eq:cons_1d_nd}\end{subequations}
where, as before, we can formally introduce the effective (dimensionless) pressure
\begin{equation}
 \tilde{\wp}(\varrho) := \wp(\varrho) + \epsilon\varsigma^2{\varrho}^{-1}.
\label{eq:effective_p}
\end{equation}
Here, it is evident that the (gradient of the) thermal displacement results in a $\mathcal{O}(\epsilon)$ or ``smaller'' correction to the pressure $\wp$, even if $\varsigma = \mathcal{O}(1)$. %in the (dimensionless) momentum equation \eqref{eq:mom_cons_1d_nd}

The linearization of the system of equations \eqref{eq:cons_1d_nd} about some state $(\varrho_i,\mathfrak{m}_i) \equiv (\varrho_i,\varrho_i v_i)$ yields
\begin{equation}
\begin{pmatrix} \varrho\\ \mathfrak{m} \end{pmatrix}_t + \begin{pmatrix} 0 & \epsilon\\ -(\epsilon\mathfrak{m}_i^2+\varsigma^2)/\varrho_i^2 + \wp'(\varrho_i)/\epsilon & 2\epsilon \mathfrak{m}_i/\varrho_i \end{pmatrix} \begin{pmatrix} \varrho\\ \mathfrak{m} \end{pmatrix}_x = 0,
\label{eq:1d_sys_lin}
\end{equation}
where
$\wp' \equiv \mathrm{d}\wp/\mathrm{d}\varrho$.
The eigenvalues $\lambda$ of the coefficient matrix in Eq.~\eqref{eq:1d_sys_lin} satisfy
\begin{equation}
\lambda^2 - (2\epsilon \mathfrak{m}_i/\varrho_i) \lambda + \epsilon (\epsilon\mathfrak{m}_i^2+\varsigma^2)/\varrho_i^2 - \wp'(\varrho_i) = 0.
\label{eq:1d_evals}
\end{equation}
For the system \eqref{eq:cons_1d_nd} to remain \emph{strictly hyperbolic} (see, \textit{e.g.}, \cite[Chap.~2]{l02}), we must require that the eigenvalues of the coefficient matrix of the linearized system \eqref{eq:1d_sys_lin} remain real and nonzero or, equivalently, that the discriminant of the polynomial \eqref{eq:1d_evals} be positive:
\begin{equation}
4\epsilon^2\mathfrak{m}_i^2/\varrho_i^2 - 4 [ \epsilon (\epsilon\mathfrak{m}_i^2+\varsigma^2)/\varrho_i^2 - \wp'(\varrho_i)] > 0,
\end{equation}
which simplifies to
\begin{equation}
 \wp'(\varrho_i) > \epsilon\varsigma^2/\varrho_i^2.
\label{eq:GN_1d_hyperbolicity}
\end{equation}
In the limit of $\varsigma \to 0^+$, the latter reduces to the usual assumption, \textit{i.e.}, $\wp'(\varrho_i) > 0$, under which the Euler system remains {strictly hyperbolic}. This assumption is, of course, equivalent to requiring that the speed of sound in the fluid remains a real number [see also Eq.~\eqref{eq:soundspd} below and the discussion thereof].

More importantly, however, the hyperbolicity condition given in Eq.~\eqref{eq:GN_1d_hyperbolicity} can be rewritten in terms of the effective pressure $\tilde\wp$ introduced in Eq.~\eqref{eq:effective_p} to yield
\begin{equation}
\tilde\wp'(\varrho_i) > 0.
\label{eq:wpt_cond}
\end{equation}
Therefore, under a redefinition of the functional form of the barotropic pressure, as in Eq.~\eqref{eq:effective_p}, the 1D lossless compressible GN fluid retains the same (hyperbolic) mathematical structure \cite{l73,W74} as the 1D inviscid compressible Euler fluid!\footnote{That is, \emph{as long as} Eq.~\eqref{eq:wpt_cond} is satisfied, a point to which we return in Section~\ref{sec:baro}.}

% ----------------------------------------------------------------
\section{Barotropic equation(s) of state}\label{sec:baro}

At this point in the discussion, we have not yet specified the equation of state, \textit{i.e.}, the relationship between the pressure $\wp$ and the remaining field variables ($\varrho$, $\bm{v}$, $\bm{\delta}$, etc.), for the lossless compressible GN fluid under consideration. For the purposes of the present work, we only consider \emph{barotropic} fluids, \textit{i.e.}, ones for which $\wp = \wp(\varrho)$ \cite{th72,OO04}. It is worth reminding the reader that the \emph{flow} of a barotropic \emph{fluid} must be barotropic, however, the converse does not necessarily have to be true.

In particular, following \cite{ccj07,ccj15}, we consider two example of barotropic equations of state. The first is the case of a perfect gas,\footnote{For concreteness, here we use Thompson's \cite[p.~79]{th72} definition of a perfect gas, namely a gas satisfying the ideal equation of state and having constant specific heats.} for which, under the above assumption of homentropic flow, it can be shown that the equation of state assumes a polytropic form \cite{s97}, \textit{i.e.}, $\wp$ is a \emph{power law} of $\varrho$ only. The second is the case of a barotropic fluid with a quadratic equation of state. For fluids in general, especially liquids, it is not always possible to determine the mathematical relationship between $\wp$ and the other thermodynamic variables \cite{mo96}, hence expanding $\wp$ to second order in $\varrho$ (at constant entropy $\eta$) about its equilibrium value is an approximation that is often made \cite{mo96,lka07}.

\subsection{Homentropic flow of a perfect gas}\label{sec:homentrop_gas_0}

As mentioned above, for the homentropic flow of a perfect gas, the equation of state assumes a (dimensionless) polytropic from:
\begin{equation}
 \wp(\varrho) = \varrho^\gamma/\gamma,
\label{eq:polytrop}
\end{equation}
where we remind the reader that the constant $\gamma(>1)$ is the {adiabatic index}, and we have used the dimensionless variables from Eq.~\eqref{eq:ndv}.

It can be verified that the equation of state \eqref{eq:polytrop} satisfies the criterion $\wp'(\varrho) > 0$ $\forall\varrho>0$, which was discussed in Section~\ref{sec:1d_sys} as necessary to ensure hyperbolicity for the 1D Euler system; and, $\wp'(\varrho) = 0 $ \emph{only} for the (trivial) vacuum state $\varrho = 0$.

\subsection{Homentropic flow with a quadratic equation of state}\label{sec:homentrop_liq}

As mentioned above, for fluids in general and liquids in particular, an equation of state can be obtained by expanding $\wp$ to second order in $\varrho$ about its equilibrium value, namely,
\begin{equation}\begin{aligned}
 \wp(\varrho) &= \wp_0 + (\varrho - 1) + \frac{B}{2A}(\varrho - 1)^2 + \cdots\\
              &= \wp_0 + (\varrho - 1) + (\beta-1)(\varrho - 1)^2 + \cdots,
\label{eq:quadrat_barotrop}
\end{aligned}\end{equation}
where $B/A$ is known as the \emph{nonlinearity parameter} \cite{b97,lka07} and is related to the \emph{coefficient of nonlinearity} $\beta$ via $\beta = 1 + B/(2A)$ \cite{lka07}. %\in (1,7)$ for nearly all (single-phase) fluids under ordinary conditions~\cite{mo96,b97}
In Eq.~\eqref{eq:quadrat_barotrop}, $\wp_0$ is the (dimensionless) pressure in the equilibrium state, and the expansion \eqref{eq:quadrat_barotrop} has been performed at constant entropy $\eta$, which is allowed because of our assumption of homentropic flow [recall Eq.~\eqref{eq:dim_entropy}]. In this work, we take Eq.~\eqref{eq:quadrat_barotrop} as exact, \textit{i.e.}, we consider a (hypothetical) fluid that satisfies it exactly for some value of $B/A$ (or, equivalently, $\beta$).

Note that $\wp_0\ne1$ because we made the pressure dimensionless using $\varrho_0c_0^2$ in Eq.~\eqref{eq:ndv}. For example, $\wp_0=\gamma^{-1}$ and $B/A = \gamma-1$ if Eq.~\eqref{eq:quadrat_barotrop} is to be the formal Taylor-series expansion of Eq.~\eqref{eq:polytrop} about $\varrho = 1$ \cite{mo96,mo97,b97,lka07,no98}. Thus, Eq.~\eqref{eq:quadrat_barotrop} is, in fact, valid for the homentropic flow of \emph{both} perfect gasses and certain barotropic fluids.

It can be verified that the equation of state \eqref{eq:quadrat_barotrop} satisfies the criterion $\wp'(\varrho) > 0$ if and only if $\varrho > (2\beta-3)/[2(\beta-1)]$. Notice that $(2\beta-3)/[2(\beta-1)] \le 0$ for $\beta \in (1,3/2]$, in which case the inequality $\varrho > (2\beta-3)/[2(\beta-1)]$ is automatically satisfied by virtue of the fact that the density is always positive, \textit{i.e.}, $\varrho > 0$. Indeed, when Eq.~\eqref{eq:quadrat_barotrop} is the formal Taylor-series expansion of Eq.~\eqref{eq:polytrop} about $\varrho = 1$, it follows that $\beta \in (1,4/3]$ based on the restrictions on the value of the adiabatic index $\gamma$ for perfect gases \cite{j16}. Therefore, even though the equation of state \eqref{eq:quadrat_barotrop} maintains the hyperbolicity of the Euler system for perfect gases, for the (hypothetical) GN liquids, for which we have assumed this equation of state to be exact, we must impose the further restriction that $\beta \in (1,3/2]$.\footnote{It is interesting to note that the critical value $\beta = 3/2$ is also significant for other types of acoustic models \cite{kmjc11,jncw12}, though the implications of crossing this value are quite different.}  Values of $\beta$ as high as $\approx7$ \cite[p.~582]{mo96} and $\approx10$ \cite[Table 8.4]{lka07} have been measured, hence this restriction is nontrivial.

\subsection{Effective pressure of a barotropic GN fluid in 1D homentropic flow}\label{sec:hyperbolicity}

\begin{figure}[htp]
\begin{center}
  \includegraphics[width=\textwidth]{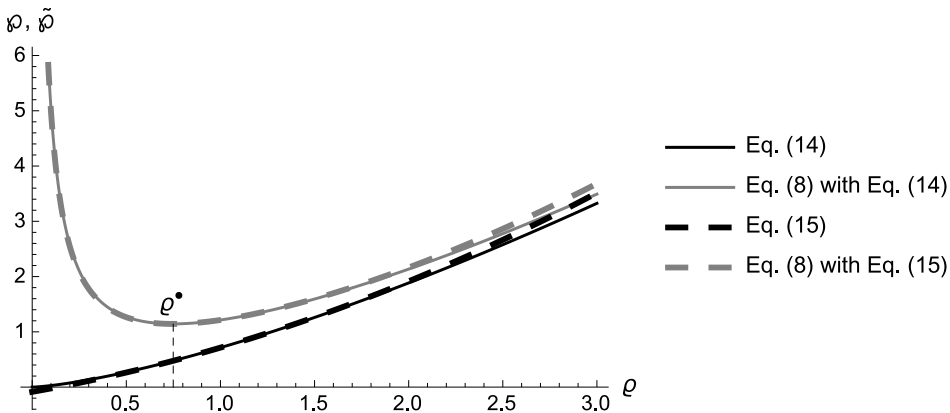}\\
  \caption{Illustration of the pressure--density relationship for the barotropic equations of states \eqref{eq:polytrop} (solid black curve) and \eqref{eq:quadrat_barotrop} (dashed black curve); the corresponding effective pressures $\tilde{\wp}$, given by Eq.~\eqref{eq:effective_p}, are shown as gray curves [solid for $\tilde{\wp}$ based on Eq.~\eqref{eq:polytrop} and dashed for $\tilde{\wp}$ based on Eq.~\eqref{eq:quadrat_barotrop}].\label{fig:pressure}}
\end{center}
\end{figure}

Figure~\ref{fig:pressure} illustrates the polytropic gas law, Eq.~\eqref{eq:polytrop}, its quadratic approximation, Eq.~\eqref{eq:quadrat_barotrop}, and the corresponding effective pressures, as defined through Eq.~\eqref{eq:effective_p}. For the purposes of Fig.~\ref{fig:pressure}, Eq.~\eqref{eq:quadrat_barotrop} is taken to be the approximation of Eq.~\eqref{eq:polytrop}, hence $\beta = 1 + (\gamma-1)/2 = (\gamma + 1)/2$. Furthermore, $\gamma = 1.4$ ($\Rightarrow$ air at 20$^\circ$C) and $\epsilon\varsigma^2 = 0.5$ are chosen for this illustration. Although $\epsilon\varsigma^2 = 0.5$ might be considered ``large,'' it leads to a better visual comparison between the thermodynamic and effective pressures, which is the purpose of Fig.~\ref{fig:pressure}.

It is clear from Fig.~\ref{fig:pressure} that the quadratic approximation [Eq.~\eqref{eq:quadrat_barotrop}] of the polytropic law [Eq.~\eqref{eq:polytrop}] is quite good for air at room temperature, across a range of densities. Of course, for $\varrho \gg 1$, we expect the quadratic approximation to become worse; indeed, the discrepancy between the solid and dashed curves is becoming noticeable near $\varrho = 3$ in Fig.~\ref{fig:pressure}. The most prominent feature of Fig.~\ref{fig:pressure}, however, is the singular behavior of the effective GN fluid pressure near the vacuum state ($\varrho = 0$). Whereas for an Euler fluid, both Eqs.~\eqref{eq:polytrop} and \eqref{eq:quadrat_barotrop} predict a finite value for $\wp(0)$, namely $0$ and $\wp_0 + \beta - 2$, respectively, the effective pressure $\tilde{\wp}$ given by Eq.~\eqref{eq:effective_p} blows up, \textit{i.e.}, $\tilde{\wp}(\varrho) \to +\infty$ as $\varrho\to0^+$.

At first glance, the latter observation might suggest that the most prominent difference in the acoustic behavior of classical (Euler) and nonclassical (GN) lossless compressible fluids might be at low densities, and this could be a regime wherein they can be experimentally differentiated. [From Eq.~\eqref{eq:effective_p}, one can show that at higher densities the difference diminishes and $\tilde{\wp} - \wp = \mathcal{O}(\varrho^{-1})$ as $\varrho \to \infty$.] However, the low density regime is problematic as, just from a visual inspection of the gray curves in Fig.~\ref{fig:pressure}, it is evident that $\tilde{\wp}'(\varrho)$ changes sign for some $\varrho \in (0,1)$ (for the chosen values of $\gamma$, $\beta$, $\epsilon$ and $\varsigma$). When we verify the condition \eqref{eq:GN_1d_hyperbolicity}, which is necessary to maintain the hyperbolicity of the 1D governing system of equations, we find that $\tilde{\wp}'(\varrho)$ changes sign at
$\varrho^\bullet = \left(\epsilon \varsigma ^2 \right)^{1/(\gamma+1)}$ for a perfect gas and at the real root\footnote{The real root can be found analytically but the expression is too lengthy to list here. The reader is encouraged to compute it using, \textit{e.g.}, {\sc Mathematica}, if needed.} of $2(\beta-1)(\varrho^\bullet)^3 - (2\beta-3)(\varrho^\bullet)^2 - \epsilon\varsigma^2=0$ for a barotropic liquid with a quadratic equation of state. Although it is expected that $\epsilon \varsigma ^2\ll1$ \cite{js06}, $\varrho^\bullet \to 0$ or $(2\beta-3)/[2(\beta-1)]$ (the bounds derived in Sections \ref{sec:homentrop_gas_0} and \ref{sec:homentrop_liq}, respectively) only in the \emph{limit} $\epsilon\varsigma^2\to0$. Hence, the issue persists for any $\epsilon\varsigma^2 \ne 0$, and $\varrho^\bullet$ could be interpreted as a lower bound on the applicability of this (nonclassical) GN theory. In other words, this GN theory appears unsuitable at low gas densities. These restrictions are, of course, related to the more specific ones derived by Jordan and Straughan \cite[Section 3]{js06} between the dimensional GN parameter $m$ and the state of the GN gas ahead of an acceleration wave.

% ----------------------------------------------------------------
\section{Model equations of nonlinear acoustics in GN fluids}\label{sec:model_eqs}

In this section, following \cite[Section~2]{js06}, we discuss the reduction of the governing systems of equations from Section~\ref{sec:intro} using a potential function for the acoustic field and the scalar thermal displacement, namely $\phi$ and $\alpha$ such that $\bm{v} = \bm{\nabla}\phi$ and $\bm{\delta} = \bm{\nabla}\alpha$, respectively. For the acoustic field, we are allowed to introduce a scalar potential since we have assumed that $\bm{\nabla}\bm{\times}\bm{v} = \bm{0}$, \textit{i.e.}, that the flow is irrotational. Meanwhile, $\bm{\delta}$ is, \emph{by definition}, the gradient of the (scalar) thermal displacement $\alpha$ \cite{gn95,js06}.

Due to the assumption of a barotropic equation of state, it is possible to introduce the (dimensionless) specific enthalpy function (see also \cite{mo96,mh97,no98}) through the definition
\begin{equation}
 h(\varrho) := h_0 + \int_{1}^\varrho \frac{1}{\hat\varrho}\frac{\rd \wp}{\rd \hat\varrho} \,\rd \hat\varrho,
\label{eq:enthalpy}
\end{equation}
where $\hat\varrho$ is a ``dummy'' integration variable and $h_0 \equiv h(1)$ is the (dimensionless) specific enthalpy of the fluid in its equilibrium state. Using the chain rule of differentiation, one can then show that
\begin{equation}
 h_t = \frac{\mathrm{d}h}{\mathrm{d}\varrho} \varrho_t = \frac{1}{\varrho}\frac{\rd\wp}{\rd\varrho } \varrho_t,\qquad
\bm{\nabla} h = \frac{\rd h}{\rd\varrho}\bm{\nabla} \varrho = \frac{1}{\varrho}\frac{\rd\wp}{\rd\varrho } \bm{\nabla} \varrho = \frac{1}{\varrho}\bm{\nabla} \wp.
\label{eq:h_derivs}
\end{equation}
At this point in the derivation, we introduce the (dimensionless) local speed of sound (see, \textit{e.g.}, \cite{th72,lka07}):
\begin{equation}
 c^2 = \frac{\rd\wp}{\rd\varrho} = \varrho\frac{\rd h}{\rd\varrho},
\label{eq:soundspd}
\end{equation}
which is a thermodynamic variable. Note that the derivative in Eq.~\eqref{eq:soundspd} is, in fact, a partial derivate taken at fixed $\eta$, however, due our assumption of homentropic flow [recall Eq.~\eqref{eq:dim_entropy}, which implies $\eta = const.$ everywhere], it becomes a total derivative.

Using Eqs.~\eqref{eq:h_derivs} and \eqref{eq:soundspd} together with $\bm{v}=\bm{\nabla}\phi$, the conservation of mass equation \eqref{eq:mass_cons_nd} can be recast as
\begin{equation}
 %h_t + \epsilon \bm{v}\bm{\cdot} \bm{\nabla} h + \epsilon c^2 \bm{\nabla}\bm{\cdot} \bm{v} =
 h_t + \epsilon \bm{\nabla}\phi\bm{\cdot} \bm{\nabla} h + \epsilon c^2 \bm{\nabla}^2\phi = 0.
\label{eq:h_contin} %eq:h_contin_poten
\end{equation}
Meanwhile, rewriting the conservation of momentum equation \eqref{eq:mom_cons_nd} in terms of $\phi$ and $\alpha$ and employing Eq.~\eqref{eq:h_derivs}$_2$, we obtain
\begin{equation}
\bm{\nabla}\phi_t +\tfrac{1}{2}\epsilon\bm{\nabla} |\bm{\nabla}\phi|^2  + \epsilon^{-1}\bm{\nabla}h = -\tfrac{1}{2}\varsigma^2 \bm{\nabla}|\bm{\nabla}\alpha|^2.
\label{eq:mom_before_int}
\end{equation}
Integrating Eq.~\eqref{eq:mom_before_int} over space and enforcing the equilibrium conditions at $t=0$, a relation between $h$, $\phi$ and $\alpha$ emerges:
\begin{equation}
h = h_0 + \tfrac{1}{2}\epsilon\varsigma^2 - \epsilon\left(\phi_t + \tfrac{1}{2} \epsilon |\bm{\nabla} \phi|^2 + \tfrac{1}{2}\varsigma^2 |\bm{\nabla}\alpha|^2 \right),
\label{eq:phi_h}
\end{equation}
which generalizes the so-called \emph{Cauchy--Lagrange integral} \cite{s97} for the Euler equations.
For convenience, let us define $\mathfrak{h}_0 := h_0 + \tfrac{1}{2}\epsilon\varsigma^2$.
%For convenience, let us define the (dimensionless) enthalpy in the equilibrium state to include the magnitude of the thermal displacement gradient, \textit{i.e.}, $\mathfrak{h}_0 := h_0 + \tfrac{1}{2}\epsilon\varsigma^2$, where recall that $|\bm{\nabla}\alpha|^2 \equiv |\bm{\delta}|^2 = 1$ in the fluid's initial equilibrium state under the nondimensionalization scheme in Section~\ref{sec:nd}.

Finally, introducing Eq.~\eqref{eq:phi_h} into Eq.~\eqref{eq:h_contin} yields
\begin{multline}
 \phi_{tt} + 2\epsilon \bm{\nabla} \phi\bm{\cdot}\bm{\nabla}\phi_t + \tfrac{1}{2} \epsilon^2 \bm{\nabla}\phi\bm{\cdot} \bm{\nabla} |\bm{\nabla} \phi|^2 = c^2 \bm{\nabla}^2\phi \\
 - \varsigma^2 \left(\bm{\nabla}\alpha\bm{\cdot}\bm{\nabla}\alpha_t +  \tfrac{1}{2}\epsilon\bm{\nabla}\phi\bm{\cdot} \bm{\nabla} |\bm{\nabla}\alpha|^2\right),
\label{eq:moment_phi}
\end{multline}
which is an \emph{exact} expression of the conservation of momentum for a lossless compressible GN fluid in terms of the scalars $\phi$ and $\alpha$.
Equation~\eqref{eq:moment_phi} is not closed because, at this point, we have not specified $c$ in terms of $\phi$ and $\alpha$. In addition, Eq.~\eqref{eq:moment_phi} is still coupled to Eq.~\eqref{eq:delta_cons_nd} through the thermal displacement gradient $\bm{\nabla}\alpha \equiv \bm{\delta}$. In terms of $\alpha$ and $h$, Eq.~\eqref{eq:delta_cons_nd} can be rewritten as
\begin{equation}
%\varrho\bm{\nabla}\bm{\cdot}\bm{\delta} + \bm{\nabla}\varrho \bm{\cdot}\bm{\delta} =
c^2\bm{\nabla}^2\alpha + \bm{\nabla}h\bm{\cdot}\bm{\nabla}\alpha = 0
\end{equation}
or, upon introducing Eq.~\eqref{eq:phi_h} into the latter,
\begin{equation}
c^2\bm{\nabla}^2\alpha + \epsilon\bm{\nabla}\left(\phi_t + \tfrac{1}{2} \epsilon |\bm{\nabla} \phi|^2 + \tfrac{1}{2}\varsigma^2 |\bm{\nabla}\alpha|^2 \right)\bm{\cdot}\bm{\nabla}\alpha
= 0.
\label{eq:GN_alpha_phi}
\end{equation}

Next, we consider two cases in which $c$ can be written explicitly in terms of $\phi$: homentropic flow of (\textit{i}) a perfect gas and (\textit{ii}) a fluid with a quadratic equation of state.

\subsection{Homentropic flow of a perfect gas}\label{sec:homentrop_gas}
Referring to \cite{ccj07} for the details, it can be shown that, in this case,
\begin{equation}
 h(\varrho) = \frac{1}{\gamma -1} \varrho^{\gamma-1},\qquad h_0 = \frac{1}{\gamma -1}
\label{eq:h-rho_poly}
\end{equation}
and
\begin{equation}
 c^2(h) = (\gamma-1)h \equiv 1 + (\gamma-1)(h-h_0).
\label{eq:csqofh}
\end{equation}
Finally, using Eq.~\eqref{eq:phi_h} to eliminate $h$, Eq.~\eqref{eq:csqofh} furnishes the closure relation
\begin{equation}%\begin{aligned}
 c^2(\phi,\alpha) %&= (\gamma-1)\left[\mathfrak{h}_0 - \epsilon\left( \phi_t + \tfrac{1}{2} \epsilon |\bm{\nabla} \phi|^2 + \tfrac{1}{2}\varsigma^2 |\bm{\nabla}\alpha|^2  \right) \right]\\
  = 1 - \epsilon(\gamma-1)\left[ \phi_t + \tfrac{1}{2} \epsilon |\bm{\nabla} \phi|^2 + \tfrac{1}{2}\varsigma^2( |\bm{\nabla}\alpha|^2 - 1) \right].
%\end{aligned}
\label{eq:c2_h_polytrop}\end{equation}

\subsection{Homentropic flow with a quadratic equation of state}\label{sec:homentrop_quadrat}
Referring to \cite{ccj07} for the details, it can be shown that, in this case,
\begin{equation}
h(\varrho) = (3 - 2\beta) (\ln \varrho + \tilde\beta \varrho),\qquad h_0 = 2(\beta-1),
\label{eq:h-rho_baro}
\end{equation}
where we have defined $\tilde\beta :=  2(\beta-1)/(3-2\beta) \equiv (B/A) / (1-B/A)$ for convenience, and
\begin{equation}
c^2(h) = (3 - 2\beta)\left\{ 1 + W_0\bigl[\tilde\beta \re^{h/(3 - 2\beta)}\bigr] \right\},
\label{eq:csqofh2}
\end{equation}
where $W_0(\cdot)$ is the principal branch of the Lambert $W$-function \cite{cghjk96}, a special function function with a surprising number of applications in the physical sciences (see, \textit{e.g.}, \cite{pmj14}).
Finally, using Eq.~\eqref{eq:phi_h} to eliminate $h$, Eq.~\eqref{eq:csqofh2} furnishes the closure relation
\begin{multline}
c^2(\phi,\alpha) = (3 - 2\beta)\\ \times \Bigg[ 1 + W_0\left(\tilde\beta \exp\left\{ \frac{1}{(3 - 2\beta)}\left[\mathfrak{h}_0 - \epsilon\left( \phi_t + \tfrac{1}{2} \epsilon |\bm{\nabla} \phi|^2 + \tfrac{1}{2}\varsigma^2 |\bm{\nabla}\alpha|^2 \right) \right]\right\}\right) \Bigg].
\label{eq:c2_h_barotrop}
\end{multline}
Note that by obtaining the closed-form expression \eqref{eq:c2_h_barotrop}, we have achieved an \emph{exact} reduction of the governing equations (in terms of the scalars $\phi$ and $\alpha$) for the homentropic flow of a barotropic GN fluid with a quadratic equation of state, just as in the case of a perfect GN gas (Section~\ref{sec:homentrop_gas}).

Finally, following \cite{ccj07}, we can expand Eq.~\eqref{eq:csqofh2} in a Taylor series in $h$ about $h_0$ [motivated by the fact that, from Eq.~\eqref{eq:phi_h}, $h-h_0 = \mathcal{O}(\epsilon)$], keeping terms up to $\mathcal{O}[(h-h_0)^2]$ and using the identity $W_0(\xi \re^{\xi}) = \xi$ \cite{cghjk96}, to obtain
\begin{equation}
c^2(h) = 1 + 2(\beta-1)(h-h_0) + \mathcal{O}[(h-h_0)^2],
\end{equation}
which, using Eq.~\eqref{eq:phi_h} to eliminate $h$, becomes
\begin{equation}
c^2(\phi,\alpha) \approx 1 - 2\epsilon(\beta - 1)\left[ \phi_t + \tfrac{1}{2} \epsilon |\bm{\nabla} \phi|^2 + \tfrac{1}{2}\varsigma^2 \left(|\bm{\nabla}\alpha|^2 - 1\right) \right].
\label{eq:c2_h_barotrop_fa}
\end{equation}
Thus, by recalling that $\beta = 1 + (\gamma-1)/2$ for a perfect gas, we see that Eqs.~\eqref{eq:c2_h_barotrop_fa} and \eqref{eq:c2_h_polytrop} are equivalent, and that the sound speed in a perfect gas and in a liquid with a qudratic barotropic equation of state coincide in the weakly nonlinear regime, \textit{i.e.}, for $\epsilon \ll 1$ (see also the discussion in \cite{ccj07}).

\subsection{Weakly nonlinear model equations}\label{sec:wn_eqs}

In this section, we derive consistent weakly nonlinear (also known as ``finite amplitude'')  approximations by neglecting terms of $\mathcal{O}(\epsilon^2)$, in the spirit of Blackstock \cite{B63}, Lesser and Seebass \cite{LS68} and Crighton~\cite{C79}, who pioneered similar approaches for the thermoviscous case. In \cite{ccj07}, it was shown that the consistent weakly nonlinear approximation, which does not involve ``unnecessary'' further modifications of the nonlinear terms, results in a solution closest to the reference Euler solution for a model shock tube problem in the $\epsilon\ll1$ regime. In the context of GN theory, other weakly nonlinear model acoustic equations have also been proposed \cite{js06} based on further (approximate) manipulations of the nonlinear terms.

As noted in Section~\ref{sec:homentrop_quadrat}, the exact equations governing the homentropic flow of a perfect GN gas are identical to the approximate equations governing the homentropic flow of a barotropic GN liquid with a quadratic equation of state, in the weakly nonlinear ($\epsilon\ll1$) regime. Thus, to derive the consistent (also known as ``straightforward'' \cite{ccj07}) weakly nonlinear approximation, we drop all terms explicitly of $\mathcal{O}(\epsilon^2)$ in Eqs.~\eqref{eq:moment_phi}, \eqref{eq:GN_alpha_phi} and \eqref{eq:c2_h_barotrop_fa}, arriving at
\begin{subequations}\begin{align}
 \phi_{tt} + 2\epsilon \bm{\nabla} \phi\bm{\cdot}\bm{\nabla}\phi_t &= c^2 \bm{\nabla}^2\phi
 - \varsigma^2 \left(\bm{\nabla}\alpha\bm{\cdot}\bm{\nabla}\alpha_t +  \tfrac{1}{2}\epsilon\bm{\nabla}\phi\bm{\cdot} \bm{\nabla} |\bm{\nabla}\alpha|^2\right),\label{eq:sf_wne_mom}\\
 c^2\bm{\nabla}^2\alpha &= -\epsilon\bm{\nabla}\left(\phi_t + \tfrac{1}{2}\varsigma^2 |\bm{\nabla}\alpha|^2 \right)\bm{\cdot}\bm{\nabla}\alpha,\label{eq:sf_wne_alpha}\\
 c^2 &= 1 - 2\epsilon(\beta - 1)\left[ \phi_t + \tfrac{1}{2}\varsigma^2 \left(|\bm{\nabla}\alpha|^2 - 1\right) \right].\label{eq:sf_wne_c2}
\end{align}\label{eq:sf_wne}\end{subequations}
Clearly, Eq.~\eqref{eq:sf_wne_c2} can be introduced into Eqs.~\eqref{eq:sf_wne_mom} and \eqref{eq:sf_wne_alpha} to yield a system of two coupled partial differential equations (PDEs) for the scalars $\phi$ and $\alpha$.

At this point in the discussion, we have not specified the order of magnitude of the GN number $\varsigma$.\footnote{Since GN fluids have not yet been observed in nature, there are no representative values of the GN parameter $m$ that could be used to estimate the GN number $\varsigma$.} Jordan and Straughan \cite{js06} argued that the nonclassical effects must necessarily be small, therefore it may be appropriate to take $\varsigma = \mathcal{O}(\epsilon)$. In this regime, Eqs.~\eqref{eq:sf_wne} simplify further, since we are neglecting all terms of $\mathcal{O}(\epsilon^2)$:
\begin{subequations}\begin{align}
 \phi_{tt} + 2\epsilon \bm{\nabla} \phi\bm{\cdot}\bm{\nabla}\phi_t &= [1 - 2\epsilon(\beta - 1)\phi_t] \bm{\nabla}^2\phi,\label{eq:sf_wne_mom2}\\
 [1 - 2\epsilon(\beta - 1)\phi_t]\bm{\nabla}^2\alpha &= - \epsilon\bm{\nabla}\phi_t \bm{\cdot}\bm{\nabla}\alpha.\label{eq:sf_wne_alpha2}
\end{align}\label{eq:sf_wne2}\end{subequations}
Equations~\eqref{eq:sf_wne2} are a \emph{one-way} coupled system, meaning Eq.~\eqref{eq:sf_wne_mom2} can first be solved, independently of Eq.~\eqref{eq:sf_wne_alpha2}, for $\phi$. Then, $\alpha$ can be found by solving the steady ``advection--diffusion''  Eq.~\eqref{eq:sf_wne_alpha2} with $\phi$ given by the solution of Eq.~\eqref{eq:sf_wne_mom2}. Therefore, in the regime of $\varsigma = \mathcal{O}(\epsilon)$, the nonclassical heat conduction effects modeled by GN theory do \emph{not} affect the acoustic field, and we readily recognize Eq.~\eqref{eq:sf_wne_mom2} as the lossless version of the so-called Blackstock--Lesser--Seebass--Crighton (BLSC) equation \cite[Section~1]{jncw12} (see also \cite[Eq.~(3.5)]{ccj07}).

Since only the square of $\varsigma$ appears in Eqs.~\eqref{eq:sf_wne}, one could consider the regime $\varsigma = \mathcal{O}(\epsilon^{1/2})$, which still corresponds to a ``small'' GN number $\varsigma$, however, now the GN number is (asymptotically) larger than the Mach number $\epsilon$. In such a weakly nonlinear regime, Eqs.~\eqref{eq:sf_wne} simplify to
\begin{subequations}\begin{align}
 \phi_{tt} + 2\epsilon \bm{\nabla} \phi\bm{\cdot}\bm{\nabla}\phi_t &= [1 - 2\epsilon(\beta - 1) \phi_t] \bm{\nabla}^2\phi
 - \varsigma^2 \bm{\nabla}\alpha\bm{\cdot}\bm{\nabla}\alpha_t ,\label{eq:sf_wne_mom3}\\
 [ 1 - 2\epsilon(\beta - 1) \phi_t]\bm{\nabla}^2\alpha &= - \epsilon\bm{\nabla}\phi_t \bm{\cdot}\bm{\nabla} \alpha.\label{eq:sf_wne_alpha3}
\end{align}\label{eq:sf_wne3}\end{subequations}
This appears to be, in a sense, the ``simplest'' straightforward weakly nonlinear model for a lossless compressible GN fluid in which there is a direct coupling between the acoustic and thermal displacement fields.

% ----------------------------------------------------------------
\section{Shock formation}\label{sec:shocks}

In this section, we present a numerical study of 1D shock formation in the homentropic flow of a lossless compressible GN fluid. In particular, we are interested in shock formation via \emph{acceleration wave blow up} (see, \textit{e.g.}, \cite{jc05,cjc06}). This type of common scenario, first considered in \cite{t57} and later extended in \cite{w76,e77,ls78}, is one of a general class of problems of discontinuity formation from smooth initial data \cite{a70,W74,fs91,ls01} with applications throughout continuum mechanics \cite{c73}, and even in social and biological systems \cite{bs14,s14}. In the acoustics context, shock formation via acceleration wave blow up is, in particular, relevant to the theory of shock tubes and resonators \cite{sh60,c64}, of which organ pipes are one example \cite{oo01}.

For the numerical study presented in this section, we use the so-called MUSCL--Hancock scheme \cite{t09} to solve the hyperbolic system of conservation laws \eqref{eq:cons_1d_nd}; for implementation details, the reader is referred to \cite{cjc06} and \cite[Appendix A]{ccj07}, while noting the typographical correction regarding \cite[Appendix A]{ccj07} given in \cite{ccj15}. The spatial grid size used is $\Delta x = 7.5\times 10^{-4}$, while the temporal step size is chosen adaptively \cite{cjc06,ccj07} to satisfy the Courant--Friedrichs--Lewy stability condition (see \cite{t09,l02}) in every computational cell at every time step. The spatial interval on which the system is solved is chosen to be large enough so that no reflections occur from the downstream boundary.

Now, let us define the dimensionless acoustic density (also known as the \emph{condensation}) $\rho\equiv \varrho -1$. Then, we consider the 1D system of conservation laws \eqref{eq:cons_1d_nd} subject to the initial conditions
\begin{equation}
\rho(x,0) = 0,\qquad v(x,0) = 0,
\label{eq:ic}
\end{equation}
and the boundary condition
\begin{equation}
\rho(0,t) = [H(t)-H(t-t_\mathrm{w})]f(t),
\label{eq:bc}
\end{equation}
where $H(\cdot)$ is the Heaviside unit step function. The two conditions in Eq.~\eqref{eq:ic} reflect the fact that the medium is initially in its equilibrium state, while Eq.~\eqref{eq:bc} describes a pulse of finite duration $t_\mathrm{w}$ (equivalently, finite width) being introduced at the $x=0$ boundary of the domain at time $t=0^+$.
Specifically, we consider a sinusoidal pulse $f(t)=\epsilon\sin(\pi t)$, which is particularly relevant to acoustics (see, \textit{e.g.}, \cite[Section~III.B]{b62}).

For the initial and boundary conditions given in Eqs.~\eqref{eq:ic} and \eqref{eq:bc}, although $\varrho$ and $v$ are initially continuous, $\varrho_t$ (and, through the coupled nature of $\varrho$ and $v$, $v_t$) suffers a jump discontinuity across the surface $x=0$. Let $x=\Sigma(t)$ be the location of this jump discontinuity at later times with $\Sigma'(t)$ being its velocity \emph{relative to the fluid} \cite{c73}. Then, we define the jump of any variable, say $\mathfrak{F}$, across $\Sigma$ as
\begin{equation}
[\![\mathfrak{F}]\!] := \mathfrak{F}^- - \mathfrak{F}^+,\qquad \mathfrak{F}^\mp := \lim_{x\to\Sigma(t)^\mp} \mathfrak{F}(x,t).
\end{equation}
If $\mathfrak{F}^- \ne \mathfrak{F}^+$ (\textit{i.e.}, $[\![\mathfrak{F}]\!]\ne0$) , then $\Sigma$ is termed a \emph{singular surface} \cite{c73}.
Since we have constructed a problem in which $[\![\varrho]\!] = [\![v]\!] = 0$ but $[\![v_t]\!] \ne 0$, the singular surface $\Sigma$ is classified as an \emph{acceleration wave} \cite{c73}.

Under the earlier assumption that the medium ahead of the wavefront $\Sigma$ is at rest, Jordan and Straughan \cite[Eqs.~(3.21) and (3.22)]{js06} found,\footnote{Note that \cite[Section 3({\it d})]{js06} uses the notation `$\alpha_x^0$', which appears to represent `$\alpha_x^+$'.} using \emph{singular surface theory}, that the jump amplitude $A(t) := [\![v_t]\!]$ obeys
\begin{equation}
  A(t) = \frac{A(0)}{1 - \epsilon a_0 A(0) t },\qquad  a_0 = \beta + \epsilon \varsigma^2, %(\varrho^+)^{-2}
\label{eq:ampl}
\end{equation}
where we have made the expression from \cite{js06} dimensionless using the variables introduced in Eq.~\eqref{eq:ndv} and, for a medium ahead at rest, used the identity $\alpha_x^+ = \delta^+ = 1/\varrho^+ = 1$, which is valid for unidirectional flow [recall Eq.~\eqref{eq:delta-rho}]. Furthermore, it can be shown that, if one of the first derivatives of a field variable suffers a jump discontinuity, then so do the rest, and all the jumps can be related to $A(t)$ \cite[Eqs.~ (3.9) and (3.19)]{js06}. Specifically, noting that $V$ in \cite{js06} denotes $\Sigma'$, which we make dimensionless via $V = c_0 V^\star$, and recalling that $v^+=0$ and $\varrho^+=1$ for a wavefront advancing into a medium at rest, \cite[Eqs.~(3.9)]{js06} become
\begin{equation}
[\![v_x]\!](t) = -\frac{1}{\Sigma'(t)}A(t),\quad [\![\varrho_t]\!](t) = \frac{\epsilon}{\Sigma'(t)}A(t),\quad [\![\varrho_x]\!](t) = -\frac{\epsilon}{[\Sigma'(t)]^2}A(t).
\label{eq:jumps}
\end{equation}
Note the appearance of $\Sigma'$ in the latter relations, which is \emph{not} the case for an acceleration wave advancing into an Euler fluid at rest (see, \textit{e.g.}, \cite[Eq.~(18)]{cjc06}). Now, from Eq.~\eqref{eq:bc}, it follows that $[\![\varrho_t]\!](0) = \epsilon\pi$, hence $A(0) = \pi \Sigma'(0)$ via Eq.~\eqref{eq:jumps}$_2$.

Similarly, upon introducing the appropriate dimensionless variables, integrating in time with $\Sigma(0)=0$ and simplifying for a medium ahead at rest, \cite[Eq.~(3.10)]{js06} gives
\begin{equation}
  \Sigma(t) = t \sqrt{1 - \epsilon \varsigma^2}. %(\varrho^+)^{-2}}.
\label{eq:sigma}
\end{equation}
Finally, from Eqs.~\eqref{eq:ampl}--\eqref{eq:sigma}, we find an expression for the slope of $\varrho(x,t)$ at the wavefront, \textit{i.e.}, at $x=\Sigma(t)$,
\begin{equation}
%[\![\varrho_x]\!](t) = -\frac{\varrho^+}{[\sqrt{1 - \epsilon \varsigma^2 (\varrho^+)^{-2}}]^2} \frac{ \pi [ \sqrt{1 - \epsilon \varsigma^2 (\varrho^+)^{-2}}]/\varrho^+}{1 - \epsilon a_0 \{ \pi [ \sqrt{1 - \epsilon \varsigma^2 (\varrho^+)^{-2}}]/\varrho^+ \} t },
 [\![\varrho_x]\!](t) = - \frac{ \epsilon\pi }{\left[1 - \epsilon a_0 \left( \pi \sqrt{1 - \epsilon \varsigma^2} \right) t \right]\sqrt{1 - \epsilon \varsigma^2}},
\label{eq:rho_slope}
\end{equation}

Since, for the problem posed above, the acceleration wave is \emph{compressive} [\textit{i.e.},  $A(0) >0$] \cite{c73}, we expect acceleration wave blow up. In this case, it is easy to see that Eq.~\eqref{eq:ampl} implies a blow up time, \textit{i.e.}, $t_\infty$ such that $\lim_{t\to t_\infty^-} A(t) = +\infty$, of
\begin{equation}
  t_\infty = \frac{1}{\epsilon a_0 A(0)}.
\label{eq:tinf}
\end{equation}
The expressions in Eqs.~\eqref{eq:ampl}, \eqref{eq:sigma}, \eqref{eq:rho_slope} and \eqref{eq:tinf} are also valid for the Euler fluid by simply taking the limit $\varsigma \to 0^+$.
Specifically, $\lim_{\varsigma\to0^+} [\![\varrho_x]\!](t) = -\epsilon\pi/(1-\epsilon a_0\pi t)$, $\lim_{\varsigma\to0^+} \Sigma(t) = t$ and $\lim_{\varsigma\to0^+} a_0 = \beta$ with $\beta = (\gamma +1)/2$ for a perfect gas, which are the corresponding expressions for an acceleration wave advancing into a perfect (Euler) gas at rest \cite{jc05,cjc06,ccj07}. Moreover, for unidirectional flow, owing to Eq.~\eqref{eq:delta-rho} and \cite[Eqs.~ (3.9) and (3.19)]{js06}, we can immediately conclude that an acceleration wave necessarily implies a \emph{temperature-rate wave} \cite{gp68,c69,m06,S11}, as was also shown in \cite{js06}.

\begin{figure}[htp]
\begin{center}
  \includegraphics[width=\textwidth]{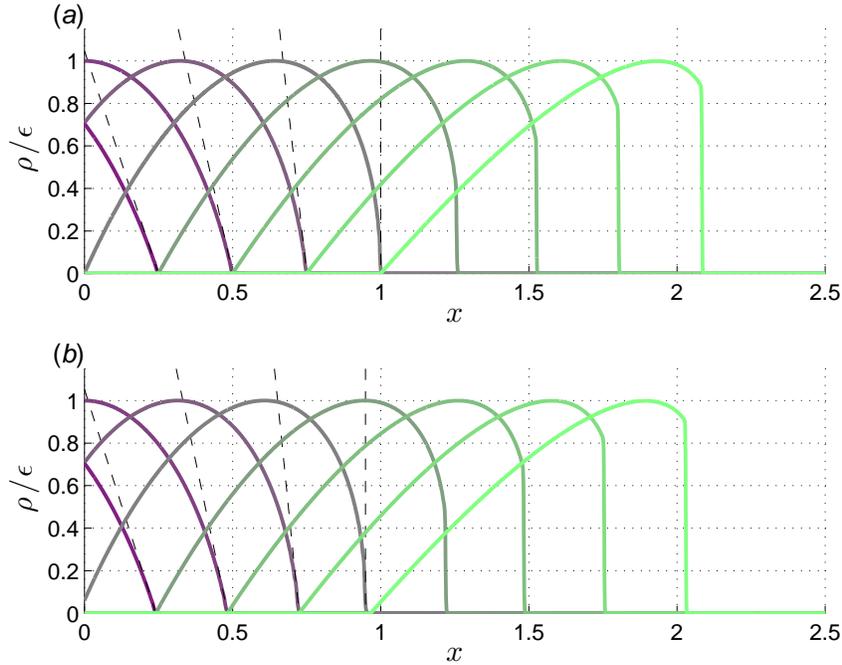}\\
      \caption{(Color online.) Shock formation via acceleration wave blow up in the hometropic flow of ({\it a}) a classical (Euler, $\varsigma = 0$) and ({\it b}) a nonclassical (GN, $\varsigma=0.5$) lossless compressible fluid. Lighter colors correspond to numerical solutions at later times; specifically, in each panel, from left to right, the solution is shown at $t=0.25$, $0.50$, $0.75$, $0.98$ [({\it b}) only], $1.00$ [({\it a}) only], $1.25$, $1.5$, $1.75$, $2.00$. The dashed lines correspond to the theoretical prediction of the slope at the wavefront, namely $\rho = [\![\varrho_x]\!](t)[x-\Sigma(t)]$, where $[\![\varrho_x]\!](t)$ and $\Sigma(t)$ are given by Eqs.~\eqref{eq:rho_slope} and \eqref{eq:sigma}, respectively. The slope is shown only \emph{prior} to blow up [\textit{i.e.}, $t\le t_\infty (\varsigma=0) \approx 1.00$ for the Euler fluid and $t\le t_\infty  (\varsigma=0.5) \approx 0.98$ for the GN fluid], at which time it becomes infinite (vertical).\label{fig:shock}}
\end{center}
\end{figure}

For the purposes of illustrating shock formation via acceleration wave blow up, let us take the medium to have similar properties to air at $20^\circ$C, \textit{i.e.}, $\gamma = 1.4$ ($\Rightarrow\beta = 1.2$), while $\epsilon = 0.26503$ and $t_\mathrm{w} = 1$. Then, for these parameter values, the predicted blow up time of the acceleration wave's amplitude is $t_{\infty} \approx 1.00$ for $\varsigma = 0$ (\textit{i.e.}, the Euler fluid). The shock formation process is illustrated in Fig.~\ref{fig:shock} for ({\it a}) $\varsigma = 0$ and ({\it b}) $\varsigma = 0.5$, which corresponds to the scaling $\varsigma = \mathcal{O}(\epsilon^{1/2})$ for $\epsilon = 0.26503$. We observe from the dashed lines in Fig.~\ref{fig:shock} that the theoretical predictions (from singular surface theory) for the locations of and slopes at the wavefront agree very well with the numerical simulations of both the classical (Euler) and nonclassical (GN) fluids for all $t\in [0,t_\infty]$. In particular, we see that shock formation occurs earlier and the wavefront moves slower in the GN fluid than in the Euler fluid [as predicted by Eqs.~\eqref{eq:tinf} and \eqref{eq:sigma}, respectively]. The numerical simulation additionally provides the full wave profile, even behind the wavefront $\Sigma$, where singular surface theory is not applicable. We observe that the classical and nonclassical wave profiles are quite similar, consistent with the assumption that the nonclassical effects are expected to be ``weak,'' \textit{i.e.}, GN theory is a small correction to the classical theory.

% ----------------------------------------------------------------
\section{Conclusion}\label{sec:conc}

In this paper, motivated by the final remarks in \cite[p.~491]{ccj07}, we re-examined the model equations of nonlinear acoustics for a class of lossless but thermally conducting nonclassical compressible fluids described by Green--Naghdi (GN) theory \cite{gn95,gn96}, in which a thermal displacement variable $\alpha$ is introduced in addition to the usual field variables of the fluid (\textit{i.e.}, density $\varrho$, velocity $\bm{v}$ and specific entropy $\eta$). Specifically, the homentropic flow of GN fluids obeying a barotropic equation of state was studied; two representative barotropic equations of state were considered: (\textit{i}) the polytropic equation of state of a perfect gas and (\textit{ii}) a quadratic approximation to a general barotropic equation state (actually applicable to both gases and liquids), which was taken as exact.

A reduction in terms of a scalar acoustic potential and the scalar thermal displacement was achieved, yielding two systems of coupled nonlinear PDEs: (I) Eqs.~\eqref{eq:moment_phi}, \eqref{eq:GN_alpha_phi} and \eqref{eq:c2_h_polytrop}, which is exact for perfect GN gases; and (II) Eqs.~\eqref{eq:moment_phi}, \eqref{eq:GN_alpha_phi} and \eqref{eq:c2_h_barotrop}, which is exact for GN fluid with a quadratic barotropic equation of state. Additionally, under the assumption of a small Mach number ($\epsilon\ll1$), some consistent weakly nonlinear models were noted. If the GN number is asymptotically of the same order as the Mach number, \textit{i.e.}, $\varsigma = \mathcal{O}(\epsilon)$, then the thermal displacement's evolution does \emph{not} affect the acoustic field, and a one-way coupled systems of PDEs \eqref{eq:sf_wne2} is obtained. Alternatively, if the GN number is taken to be small but asymptotically larger than the Mach number, \textit{e.g.}, $\varsigma = \mathcal{O}(\epsilon^{1/2})$, then the acoustic and thermal fields become coupled, as evidenced by Eqs.~\eqref{eq:sf_wne3}.

In addition, the exact nonlinear acoustic equations were studied for unidirectional flows, and a significant simplification was achieved through the introduction of an \emph{effective pressure}. In this way, the system of conservation laws for mass and momentum in 1D takes an identical form for Euler (classical) and GN (nonclassical) fluids, as shown by Eqs.~\eqref{eq:cons_1d_nd} and \eqref{eq:effective_p}. Using this formalism, bounds were derived for the validity of GN theory. Specifically, unlike the Euler case, it was shown in Section~\ref{sec:hyperbolicity} that the system of conservation laws \eqref{eq:cons_1d_nd} loses hyperbolicity for $\varrho \in (0,\varrho^\bullet)$ for some $\varrho^\bullet$. Thus, GN theory appears to not be applicable at low fluid densities.

In the same 1D context of unidirectional flow, shock formation via acceleration wave blow up was examined numerically. A high-resolution Godunov-type (\textit{i.e.}, shock-capturing) finite-volume scheme was used to solve Eqs.~\eqref{eq:cons_1d_nd} numerically for the canonical problem of the injection of a sinusoidal density signal into a quiescent fluid. Jordan and Straughan's exact results \cite{js06} (based on singular surface theory) for acceleration wave blow up in this context  were thus confirmed numerically, specifically showing that blow up occurs earlier in the nonclassical fluid and the wavefront's velocity is slowed down in comparison to the classical fluid. Additionally, the numerical simulations allowed for a comparison of the full wave profiles (at \emph{and} behind the wavefront), showing that the wave profiles of the Euler and GN fluids are qualitatively similar for this model problem.

Finally, in the context of classical acoustics, it has been shown that a number of weakly nonlinear model equations are the Euler--Lagrange equations that extremize certain functionals  \cite{sw68,no98,rsgc11,ccj15,rsgc16}. Therefore, in future work, it may be of interest to determine whether the model acoustic equations presented herein, \textit{e.g.}, Eqs.~\eqref{eq:moment_phi}, \eqref{eq:GN_alpha_phi} and \eqref{eq:c2_h_polytrop} or Eqs.~\eqref{eq:moment_phi}, \eqref{eq:GN_alpha_phi} and \eqref{eq:c2_h_barotrop} [or even the model \emph{weakly} nonlinear acoustic equations, namely Eqs.~\eqref{eq:sf_wne}, Eqs.~\eqref{eq:sf_wne2} and Eqs.~\eqref{eq:sf_wne3}], possess such a \emph{variational structure}. Additionally, it may be worthwhile to consider stochastic effects on the dynamics of acceleration wave (see, \textit{e.g.}, \cite{ost99}) in nonclassical fluids.

% ----------------------------------------------------------------

\section*{Acknowledgments} The author would like to thank Dr.\ P.\ M.\ Jordan and the guest editors for the invitation to participate in this special issue and their efforts in organizing it.

% ----------------------------------------------------------------

% You may incorporate your references as follows in your main tex file.
% Using BibTex is not recommended but can be handled.

\medskip
% The data information below will be filled by AIMS editorial staff
Received  January 2016; revised January 2016.
\medskip

\end{document}